\documentclass{article}
\usepackage{authblk}
\usepackage{amsmath}
\usepackage{amssymb}
\usepackage{graphicx}

\def\be{\begin{equation}}
\def\ee{\end{equation}}

\def\XXint#1#2#3{{\setbox0=\hbox{$#1{#2#3}{\int}$}
     \vcenter{\hbox{$#2#3$}}\kern-.5\wd0}}

\title{Some comments on the electrostatic forces between circular electrodes}

\author[1]{Francesco Maccarrone}
\author[1,2]{Giampiero Paffuti}
\affil[1]{Dipartimento di Fisica - Universit\'a di Pisa,  Largo Pontecorvo 7, Pisa, Italy}
\affil[2]{INFN sezione di Pisa, L.go Pontecorvo 3 Ed.  C, I-56127 Pisa, Italy}

\begin{document}
\maketitle
\begin{abstract}
We study the force between two circular electrodes in  different configurations. A formula analogous to Kelvin's formula for the spheres
is given in the case of equal disks held at the same potential and when one plate is earthed.  
An expression for the force at short distance between two arbitrarily charged disks is found: the generic case shows a logarithmic repulsive force,  also for disks carrying charges of opposite sign. Some numerical computations support the results. A classification for the possible behaviors of the force is proposed on the basis of a decomposition of the capacitance matrix. It is shown that the forces depend strongly on the dimensionality of the contact zone between the conductors.
The analysis is supported by a numerical computation carried for the case of two disks of different radii.
\end{abstract}

\section{Introduction}

The  knowledge of the  electrostatic force between two  charged conductors has a theoretical and practical importance. Its exact calculation  is possible through the capacitance coefficients \cite{maxw}, whose analytical value is known only  for few selected geometries and arrangements of the electrodes. 

For large distances  between the charged bodies the problem can be understood in terms of the properties of the isolated conductors \cite{mac} and, in principle, it is solved. For short distances, the situation is more complex.  A physical argument   rests on the fact that when the two conductors touch, a new system is formed. Some unexpected properties arise in this regime. For example, two spheres of like charges almost always attract, as shown by J.Lekner \cite{lek1}.

A particularly intriguing problem is the following: which  is the force between two equal conductors at the same potential?
This problem was solved in a classical work by Lord Kelvin\cite{kelvin} for the case of two spheres of equal radius $a$ finding a repulsive finite force at short distances:
\be F = \frac{Q^2}{a^2} \dfrac{4\log(2)-1}{24\, (\log(2))^2}\label{intro1.3}\ee
Another classical case studied by Kelvin is a couple of spheres when one of the two is earthed.
These problems have been recently revisited and generalized to spheres of different radii by J.Lekner\cite{lek1}, who give an instructive study of different physical situations.

In this work we study the same problem for the case of two coaxial disks of radius $a$, when the distance $\ell$  between them goes to zero. For two disks with fixed charges $Q_1, Q_2$ we find a {\em logarithmically divergent} force at short distances, the only exception
being $Q_2=-Q_1$, when the force is attractive and constant. Moreover a curious, and potentially interesting, behaviour is observed
in this near regime when the disks carry charges of different sign and of different absolute value. We found that the interaction cancels at a given distance and become repulsive for smaller gaps. In a way, the behaviour is the counterpart of that reported for two spheres. Owing to the redistribution of charges on the surface of the disks the interaction pass from attraction to repulsion for total charges of different sign. Surprisingly, this situation appears also for two disks with charges of the same sign if the ratio of the two charge is small, $Q_1/Q_2 \lesssim 0.16$.

With conductors we can consider configurations different from the standard case of two bodies of given charges, i.e. two conductors at fixed potentials and one conductor with charge $Q$ while the other is held at fixed potential. These two cases cover the Kelvin's configurations. We will give below  the small distance behaviour of the forces in these cases, for two disks.

This paper is organized as follows. In section~\ref{sezmethod} we present the general framework and give a formal expression for the forces. In section~\ref{sez2}, a simple analysis of an integral equation provide us the needed asymptotic terms in the capacitance matrix. We complete the analysis and give the explicit expression for forces  between the disks in different cases. Particularly, we show that disks charged with different charges  of opposite sign repel each other for an appropriate reduction of the inter-electrode distance. The calculated distribution of surface charge in this case gives a physical insight in this curious result. 
In section~\ref{sezgen} we propose a generalization to the case of arbitrary conductors, based on a particular decomposition of the capacitance matrix. With this method the role of the dimensionality of the contact zone is emphasized and the corresponding different behaviour of forces is naturally explained. This approach is checked in the case of two disks with different radii, where a constant force arises at short distances.

\section{Method\label{sezmethod}}
Charges and potentials for two electrodes are linearly related
\be Q_i= \sum_j C_{ij} V_j\,;\quad V_i = \sum_j M_{ij} Q_j \label{intro1.2}\ee
$C_{ij}$ are the elements the symmetric capacitance matrix and its inverse, with elements $M_{ij}$, is the potential matrix.
The  electrostatic energy of the system is given by the quadratic forms
\be W = \frac12 \sum_{i,j} C_{ij} V_i V_j = \frac12 \sum_{ij} M_{ij} Q_i Q_j\,.\label{intro1.1}\ee
In the general case, all the quantities in  both quadratic forms of \eqref{intro1.1} depend on the distance $\ell$ between the conductors.  On the other hand, the choice of the first or the other form is convenient when conductors at fixed potentials or at fixed charges are studied.
The force between the conductors  is found by derivation of energy. For example, if the two disks have fixed charges, $Q_1$ and $Q_2$ respectively the force in the axial direction is
\be F = - \frac12\sum_{i,j} Q_i Q_j \frac{\partial}{\partial \ell}M_{ij}
\label{intro1.3}\ee

All the previous formulas simplify for equal conductors, e.g. equal disks, because  $C_{11}= C_{22}$ and $M_{11}= M_{22}$. For equal conductors, it is convenient to distinguish in the capacitance matrix the usual relative capacitance, given in this case by $C = (C_{11}- C_{12})/2$ and the symmetric combination $C_g = (C_{11}+ C_{12})$ which enters
in the computation of the energy for fixed potentials. In terms of these quantity
\be C_{11} = C + \frac12 C_g\,;\qquad C_{12} = - C + \frac12 C_g \label{decompC}\ee

Let us consider the electrodes at fixed charges. 
The matrix $M_{ij}$ is calculated inverting $C_{ij}$ and substituting in \eqref{intro1.1} we obtain for the energy
\be W = \frac14(Q_1+Q_2)^2\,\frac{1}{C_g} + \frac18(Q_1-Q_2)^2\frac1{C} \label{energia}\ee
The force is given by
\be F(Q_1,Q_2,\kappa) = - \dfrac{(Q_1+Q_2)^2}{4}\frac{\partial}{\partial\ell}
\frac1{C_g} 
- \dfrac{(Q_1-Q_2)^2}{8}\frac{\partial }{\partial\ell}
\frac1{C}
\label{forza2}\ee
 where $\kappa=\ell/a$ is called aspect ratio of the two-disks system.

The rationale behind the decomposition \eqref{decompC} lies in the isolation of the divergent  behaviour as $\ell\to 0$.
In the general case \cite{paf}
\be \lim_{\ell\to 0} (C_{11} + C_{22} + 2 C_{12} ) = C_T \label{intro1.6}\ee
where $C_T$ is the capacitance of the conductor obtained when the two separated bodies touch. For equal conductors $C_g\to C_T/2$ which is a finite quantity. Let us note that $C_g$ is the only term which enter in the computation of the force for equal charges.

It is well known that the expression \eqref{forza2} is valid in every circumstance, for equal conductors, depending only on Coulomb's law, but the dependence on $\ell$ is hidden also in the charges when the bodies are held at fixed potential. For completeness let us shortly review the case of equal potential $V_1 = V_2 = V$. The charges are equal by symmetry $Q_1 = Q_2 = Q_V$ and from
\eqref{forza2}
\be F_V = F(Q_V, Q_V,\kappa) = - Q_V^2 \frac{\partial}{\partial\ell}
\frac1{C_g}  = \frac{Q_V^2}{C_g^2} \frac{\partial}{\partial\ell} C_g
\label{forzaV}\ee
By noticing that from \eqref{intro1.2} in this case $Q_V = C_g V$ we can also write
\be F_V =  V^2 \frac{\partial}{\partial\ell} C_g
\label{forzaV2}\ee
This result can be directly obtained by expressing the electrostatic energy $W$ in terms of $V$ and taking
$ F = + \partial_\ell W$. The plus sign is due to the energy supplied by the voltage source, as explained in textbooks\cite{LL}.
Using the decomposition \eqref{decompC} the general case of two different potentials $V_1, V_2$ give rise to a force
\be F_{V_1,V_2} = {\overline V}^2\frac{\partial}{\partial\ell} C_g + \frac{\Delta V^2}2 \frac{\partial}{\partial\ell} C 
\label{forzav1v2}\ee
with $\overline V = (V_1+V_2)/2$ and $\Delta V = V_1-V_2$.

A third interesting  configuration is one conductor with charge $Q_1$, fixed, and the second one held at potential 0.
To give the explicit dependence of the force on $\ell$ we can follow the procedure used above. As $V_2=0$ the conductor 1 is at potential $V_1 = Q_1/C_{11}$ while conductor 2 has a charge $Q_2 = C_{21} V_1 = Q_1 C_{12}/C_{11}$. Substitution in \eqref{forza2} gives, after expressing $C_g$ and $C$ in terms of $C_{11}$ and $C_{12}$:
\be F_E = -\frac12 Q_1^2 \frac{\partial}{\partial\ell}\frac{1}{C_{11}} \label{forzaM}\ee 
The same result can be obtained more easily from \eqref{intro1.1}, but this derivation has the merit of eliminating any doubt about signs. Expressions \eqref{forzaV2} and \eqref{forzaM} apply to Kelvin's configurations.

In summary, the knowledge of the force between the two conductors is equivalent to that of the dependence of the coefficients of capacitance and of potential, forming matrix $\bf{C}$ and $\bf{M}$. Unfortunately, their analytic forms is limited to few cases. On the contrary, some concluding remarks can be drawn from the asymptotic behaviour of these parameters both in the far limit and in the near one. Elsewhere \cite{mac} we treated the large distance behaviour  of $C_{ij}$ and $M_{ij}$ and in this limit the forces can be easily obtained by general formulas involving the self capacitances and other intrinsic parameters of the conductors.  
Our aim is to get an explicit formula for the forces in the case of two coaxial disks of radius $a$ separated by a distance $\ell$ in the region $\ell\ll a$. Following \eqref{forza2} this purpose requires the knowledge of the behaviour of the relative capacitance $C$ and of $C_g$ in the near limit ($\kappa \to 0$) and this is the content of the following section.

\section{The capacitor with two equal disks\label{sez2}}

It is known\cite{Love,nick,sne,Carlson,paf} that the general electrostatics problem for two coaxial disks {\it of radius $a$ and distance $\ell$,} can be reduced to the solution of a pair of integral equations
\be
\begin{split}
V_1 &= F_1(t) +\int_0^1 K(t,z;\kappa) F_2(z)\,dz\,;\\ 
V_2 &= F_2(t) +  \int_0^1 K(t,z;\kappa) F_1(z)\,dz
\end{split}
\label{eqaccoppiateug1}
\ee
where $\kappa = \ell/a$ and
\be K(t,z;\kappa) = \frac{\kappa}{\pi} \left(\dfrac1{(z-t)^2 + \kappa^2}
+ \dfrac1{(z + t)^2 + \kappa^2}\right)\,.
\label{kernelsing}\ee
The charges on the disks are shown to be
\be Q_1 = a\,\frac2\pi \int_0^1 F_1(t)\,dt \,, \qquad Q_2 = a\,\frac2\pi \int_0^1 F_2(t)\,dt\,.\label{cariche1}\ee
The coefficients $C_{ij}$ can be in principle computed by solving the system \eqref{eqaccoppiateug1}, and their linear combinations $C_g$ and $C$.

$C$ can  be numerically calculated  with very high precision \cite{Norgren,fg} from \eqref{eqaccoppiateug1} but as we need
only its asymptotic expansion we give directly the known final result, for $\kappa\to 0$:
\be
C \mathop{\sim}_{\kappa\to 0}
 a\left\{ \frac{1}{4\kappa} + \frac{1}{4\pi}\left[ \log\left(16\pi \frac 1 \kappa\right) - 1\right]\right\}
 + a\left\{\frac1{16\pi^2} \kappa\left[ \left(\log(\frac{\kappa}{16\pi})\right)^2 -2 \right]\right\}
\label{corrShaw}
\ee
The first line in \eqref{corrShaw} is the classical Kirchhoff\cite{Kirchh} asymptotic expression, the second line, a subleading correction, has been computed by S.Shaw\cite{Shaw} and improved and corrected in\cite{Wigg,chew}.

\subsection{The computation of $C_g$  at small $\kappa$}
To give an estimate of $C_g = C_{11} + C_{12}$ at small $\kappa$ consider \eqref{eqaccoppiateug1} in the case $V_1 = V_2 = V$.
By symmetry $F_1=F_2$ and defining the normalized function $F_1(t) = V g(t)$
we have to solve
\be 1 = g(t) +\int_0^1 K(t,z;\kappa) g(z)\,dz\,. \label{cariche1g}\ee

From \eqref{cariche1} and \eqref{intro1.2}
\be C_{11} + C_{12} = \frac{2a}{\pi}\int_0^1 g(t) dt \equiv C_g\label{valc11c12}\ee
When $\ell\to 0$ two disks of radius $a$ become a single disk with radius $a$, whose self-capacitance is\cite{LL} $C_T = 2a/\pi$
From \eqref{intro1.6} it follows
\be \lim_{\ell\to 0} C_g = \frac12 C_T = \frac a\pi \label{limitecg}\ee
The leading term being constant in $\ell$ we have to go to the next order in $\kappa$ to get the force.
First of all let us show as \eqref{limitecg} follows from \eqref{cariche1g}. For $\kappa\to 0$ the kernel $K$ becomes singular, the expression \eqref{kernelsing} is essentially a sum of two truncated lorentzian distributions of width $\kappa$, and as it is well known a 
lorentzian distribution shrinks to a delta function as his width goes to zero. This means that the integral operator defined by $K$ 
tends to the identity operator. Then in this limit case \eqref{cariche1g} gives
\be g(t) \to \frac12 \label{valoreg0}\ee
Substitution in \eqref{valc11c12} gives the result \eqref{limitecg}. To compute the next order let us write $g(t) = \frac12 + h(t)$ and substitute in \eqref{valc11c12} 
\be \frac12 = \frac12 G(t;\kappa) + h(t) +  \int_0^1 K(t,z;\kappa) h(z)\,dz\,. \label{cariche1gnext}\ee
where
\be
G(t;\kappa) = \int_0^1 K(t,z;\kappa) dz  = 
\frac1\pi \left(\arctan\frac{1-t}{\kappa} + \arctan\frac{1+t}{\kappa} \right) \ee
A rough estimation  of the solution can be given by applying again the trick $K\to$~identity-operator for $\kappa\to 0$.
In this way one obtains
\be h(t) = \frac14 \left(1 - G(t;\kappa) \right) + {\cal O}(k) \label{valoregnext}\ee
This  coarse approximation is sufficient for our needs, but it is clear that ${\cal O}(k)$ terms are missed in this procedure, and this has been indicated in \eqref{valoregnext}.
Performing the integral \eqref{valc11c12} and expanding in $\kappa$  one obtains
\be C_g = a\left[ \frac1\pi + \frac{\kappa}{2\pi^2}\left(\beta - \log(\kappa)\right)\right]\label{valorecgasint}\ee
The coefficient $\beta $ embodies our ignorance of the linear term in $\kappa$ in the asymptotic expansion of $g(t)$.

\subsection{Force between the disks}

The expressions \eqref{corrShaw} and \eqref{valorecgasint}, respectively for $C$ and $C_g$ at small distances, give an expansion at small $\kappa$ of \eqref{forza2}.  The leading terms of the force between the plates carrying the charges $Q_1$ and $Q_2$ is
\be F(Q_1,Q_2,\kappa) = - \frac{(Q_1-Q_2)^2}{2 a^2}  + \frac{(Q_1+Q_2)^2}{8 a^2}(\beta -1 -\log\kappa) 
\label{forzaQris}\ee

 For every choice of the charges the force is  repulsive if the distance of the disks is small enough.  The force is logarithmic divergent {\it for $\ell\to 0$ in every case} except in the case of oppositely charged disks $Q_2 = - Q_1$ where the force is constant and attractive. This is true also when the sign of the charges is opposite so that for this situation the net force is attractive at large distance, asymptotically like two charge of opposite sign, then it vanishes at a given distance and become repulsive when the disks are closer. In this condition 
the disks charged of different sign repel each other. This unexpected behaviour is the counterpart of the attraction demonstrated in \cite{lek1} in the close approach of two spheres with like charges. In the following the effect is attributed to the emergence of likely charged zones at the edge of the disks. Naively one expects that for disks with charge
of the same sign the force is always repulsive, but also in this case a small region of stability can be formed as a function of $\ell$, as will be shown below by the numerical solution of \eqref{eqaccoppiateug1}.

We can calculate also the force between the disks in the two configurations analyzed by Kelvin for the spheres. With the disks maintained at the same potential $V$ \eqref{forzaV}  gives
\be F_V = \frac1{2\pi^2}\left(\log(\beta - 1-\log\kappa\right)\,V^2 =
\frac{Q_V^2}{2a^2}\left(\beta - 1 -\log\kappa\right)\label{forzaVris}
\ee
a logarithmic divergent {\it repulsive} force at short distances. The case of one earthed disk 
and the other at a fixed charge $Q_1$ is described by \eqref{forzaM} and the small $\kappa$ expansion gives
\be
F_E = - \frac{Q_1^2}{2a^2}
\label{forzaMris}
\ee
an attractive constant force. 

\subsection{Numerical evaluation of $\beta$}
To complete our results we have to give  the value of the constant $\beta$. The analytical computation of $\beta$ requires a second order application of perturbation theory, as exposed for $C$ in\cite{Shaw}. As the main conclusions of our work do not depend on the explicit value of $\beta$ we prefer at first to give a numerical estimation of this parameter. The numerical computation is also needed at this stage to verify our procedure and to have an idea of the applicability, i.e. we have to check if the approximation \eqref{valorecgasint} do work for physically reasonable values of the ratio $\kappa = \ell/a$.

We perform this computation as a part of a larger work\cite{fg} on the numerical resulys obtained in the study of the two-disks capacitors, and we refer to this work for more details.
We solved the system \eqref{eqaccoppiateug1} 
using a grid of $N$ points given by gaussian integration points. The maximum $N$ used was $N_{max} = 45000$, this number is mainly limited by computer memory. This problem is known to suffer of a slow convergence as $\kappa\to 0$, the convergence being poor for $N\kappa \lesssim 3$. For the lowest values of $\kappa$ we used the extrapolation procedure proposed in\cite{Norgren}, this and the quite large value of $N_{max}$  lead us to be confident on our results up to $\kappa= 10^{-5}$.

\begin{figure}[!ht]
\begin{center}
\includegraphics[width=0.75\textwidth]{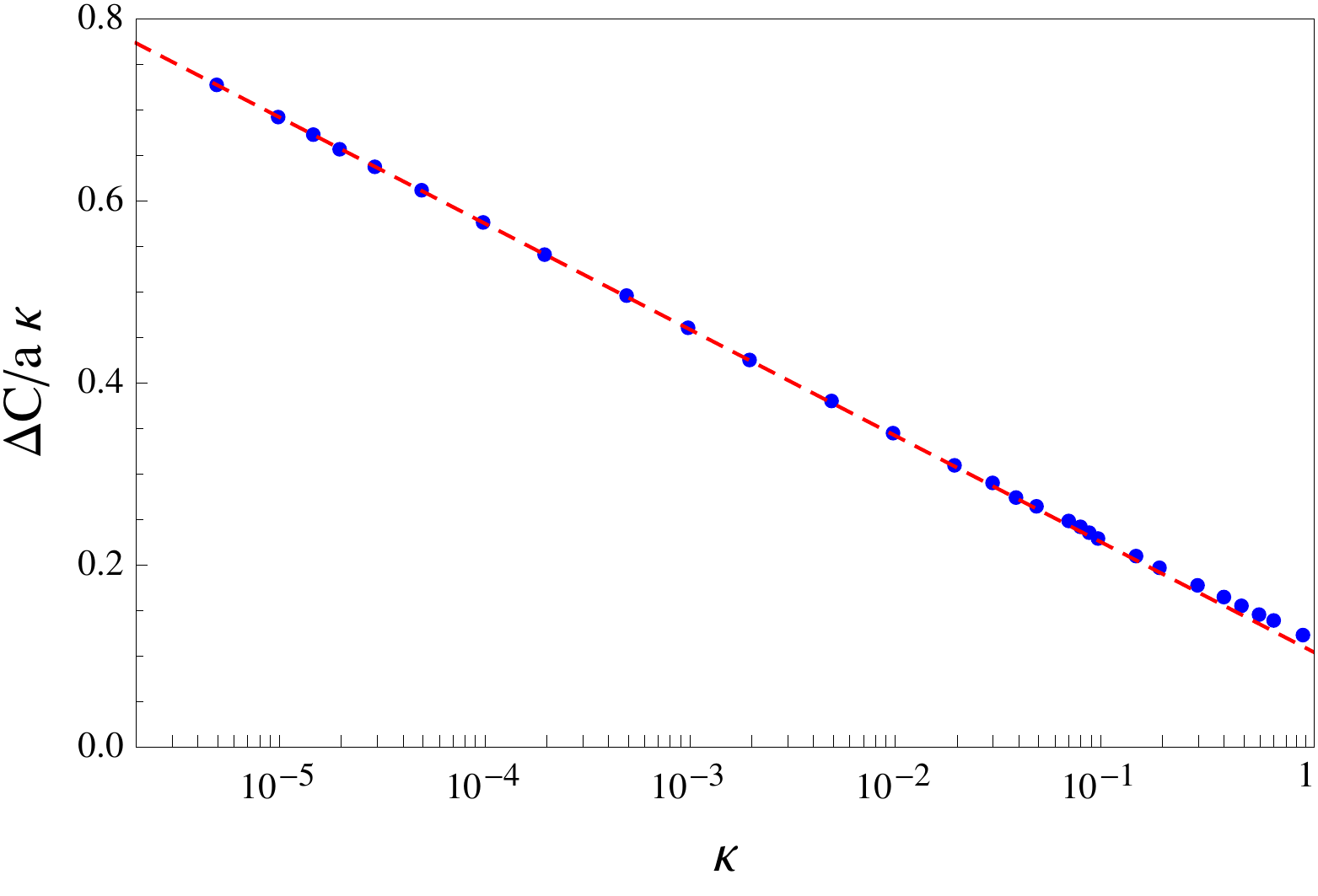}
\caption{$(C_g-a/\pi)/\kappa a$. The slope of the dashed line is $1/(2\pi^2)$ as in \eqref{valorecgasint}. The constant $\beta$ is fitted.
\label{figura1}}
\end{center}
\end{figure}

We computed separately $C_{11}$ and $C_{12}$ then we add these quantities to obtain $C_g$, this allow us to check the cancellation of all kind of divergences and the leading order result \eqref{limitecg}. Here we present a sharper result. We plot the computed values for
$\Delta C/(a \kappa)$, where
$\Delta C = C_g - a/\pi$, in the range $10^{-5} \leq \kappa \leq 1$. In a logarithmic scale for $\kappa$ the values must lie on a straight line with slope $1/(2\pi^2)$. The results are shown in figure~\ref{figura1}.

We fitted the constant $\beta$ in \eqref{valorecgasint} from the last points in figure~\ref{figura1} obtaining
\be \beta = 2.1450(2) \ee
From numerical results a reasonable conjecture is $\beta = 1 + \log(\pi)$.
From the figure it is apparent that the description \eqref{valorecgasint} is rather accurate even for $\kappa$ as large as $\kappa\sim 0.1$.

\subsection{Physical discussion of the results}

 Calculating 
the limit for $\kappa\to0$ in \eqref{energia} and taking into account the divergence of $C$ and equation \eqref{limitecg}, one finds
\be \lim_{\kappa\to0}W =  \frac{\pi}{4 a} (Q_a+Q_b)^2 \label{energiazero}\ee
which is always finite.  On the other hand, if the two disks touch  they form a single disk, with capacity $C_1 = 2 a/\pi$ and a
charge $Q = Q_a+Q_b$. The energy at equilibrium is 
\be W_0 = \frac12 C_1 Q^2 =  \frac{\pi}{4a} (Q_a+Q_b)^2\ee
i.e. there is not a change in energy due to reorganization of the charges {\it at the transition between two very close disks and two disks touching each other}.

At large distances $C_{12}\to 0, C_{11}\to C_1$ then the energy at large distances is
\be W_\infty = \frac{\pi}{4a} (Q_a^2 + Q_b^2) \label{energiazero.1}\ee
For charge of opposite sign $W_0 < W_\infty$ and, as the force is repulsive at short distances, this implies the existence of a stationary point, as already discussed in terms of the force. 

\begin{figure}[!ht]
\begin{center}
\includegraphics[width=0.75\textwidth]{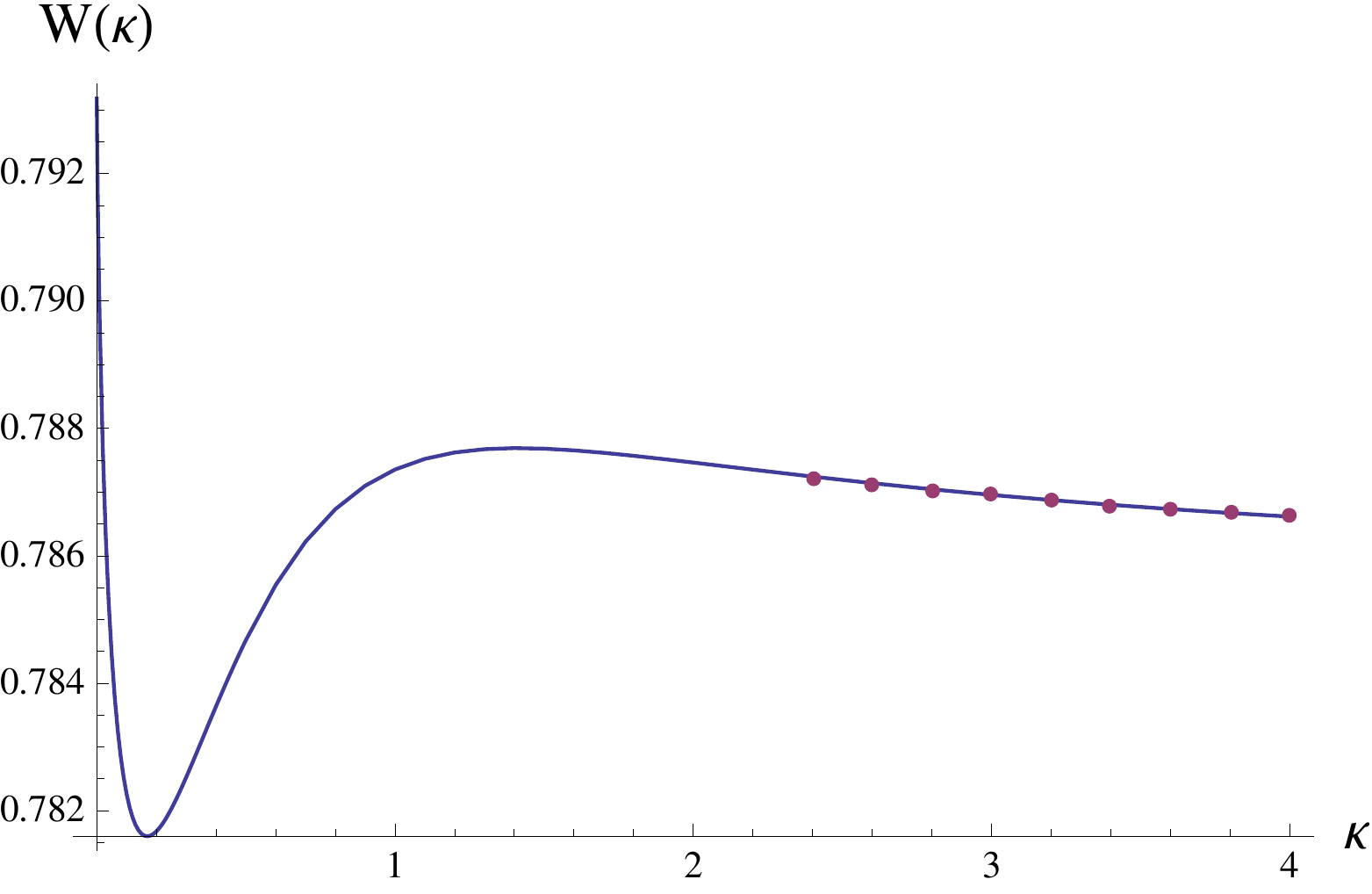}
\caption{$W$ as a function of $\kappa$ for a ratio $Q_b/Q_a = 0.005$. Points are the known asymptotic expansion at large $\kappa$.
\label{figWvsk}}
\end{center}
\end{figure}

For $Q_a, Q_b$ of the same sign instead there is no energy argument for the existence of stationary point as $W_0> W_\infty$. On the other hand, using $\beta = \log(\pi)-1$, \eqref{forzaQris} gives a zero for the force at
\[ k_0(\rho) = \pi \exp\left[- 4 \left(\frac{1-\rho}{1+\rho}\right)^2\right] \]
with $\rho = Q_b/Q_a$. If this value is in the range of validity of our approximation we expect a stationary point {\em also} for equal signs. As $k_0(0) = 0.057 < 0.1$ we expect that for small enough $\rho$ the stationary point is there. To verify this effect we computed the coefficients $C_{ij}$ numerically from \eqref{eqaccoppiateug1} and obtained the energy for varies values of $\rho$. As an example we show in figure~\ref{figWvsk} the energy $W$ in units of $Q_a^2/a$ for $\rho = +0.005$. The minimum disappears at $\rho\gtrsim 0.016$.

In the introduction we mentioned that the charge redistribution must be responsible for this unusual properties of the forces, it would be nice to verify explicitly this claim. It can be shown that the functions $F_i$ in \eqref{eqaccoppiateug1} are related to density on disks by
\[ f_i(t) = 2\pi a \int_t^1\dfrac{x}{\sqrt{x^2-t^2}}\,\sigma_i(x)\,dx \]
This is an Abel transformation which can be inverted to give \cite{Love}
\be \sigma_i(x) = \frac1{a\pi^2}\left[ \dfrac{f_i(1)}{\sqrt{1-x^2}} - \int_x^1 dt \dfrac{f_i'(t)}{\sqrt{t^2-x^2}}\right]
\label{densitasigma1}\ee
Unfortunately, \eqref{eqaccoppiateug1} in this form are not suited for a systematic study of the density at small separation as their solutions are defined at fixed potentials and the charges become enormous as $\kappa\to 0$. It would be better to write an equivalent system for fixed charges. This can be done integrating the equations in the interval $(0,1)$, using the relations \eqref{cariche1}
we have
\be
\begin{split} V_1 &= \frac{\pi}{2a} Q_1 + \int_0^1 G(s;\kappa)) F_2(s)ds\,;\\
 V_2 &= \frac{\pi}{2a} Q_2 + \int_0^1 G(s;\kappa) F_1(s) ds
\label{potcof.1}
\end{split}
\ee
where
\be G(t;\kappa) = \frac1\pi\left[\arctan\frac{1-t}{\kappa} + \arctan\frac{1+t}{\kappa}\right]\label{defG1a}\ee
Inserting back \eqref{eqaccoppiateug1} for $V_1, V_2$ we have the new system
\begin{equation}
\begin{split}
\frac{\pi}{2a} Q_1 &= F_1(t) + \int_0^1 K(t,s)F_2(s)ds- \int_0^1 G(s;\kappa)) F_2(s)ds\\
\frac{\pi}{2a} Q_2 &= F_2(t) + \int_0^1 K(t,s)F_1(s)ds -\int_0^1 G(s;\kappa) F_1(s) ds\end{split}\label{potcof2}
\end{equation}
The coefficients $M_{ij}$ can be computed directly from the solutions of this equations. Performing the transformation 
\eqref{densitasigma1} to the solutions we can easily get the density. The numerical computation confirms the claim. As an example we show
the density distributions  for two disks with charges $q_1 = +1, q_2 = -0.1$ at distance $\kappa = 0.005$.
\begin{figure}[!ht]
\begin{center}
\includegraphics[width=0.75\textwidth]{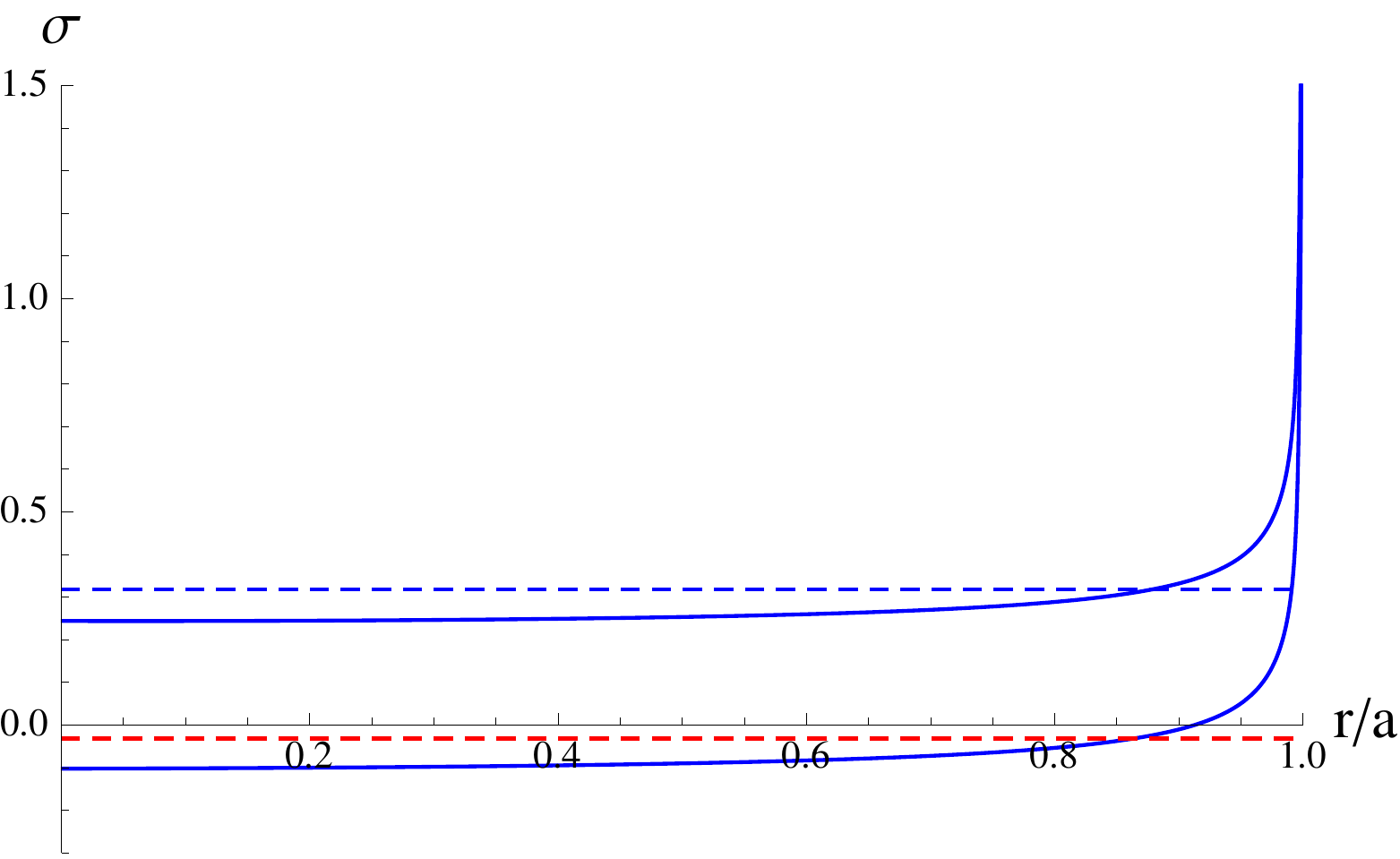}
\caption{Density distribution for two disks. The charges are $q_1 = +1, q_2 = -0.1$ and the distance $\kappa = 0.005$.
Dashed lines represents a constant charge distribution and is plotted to help the reader.
\label{densita}}
\end{center}
\end{figure}
From figure~\ref{densita} it is apparent that the negatively charged disk lower its charge in the bulk and produce an high {\em positive} density at the border, this charge repel the analogous charge on the upper disk and is responsible for the repulsive force.

On physical grounds a divergent force between macroscopic bodies is difficult to accept, then it is natural to ask for 
mechanisms that can inhibit the logarithmic grow. An obvious cutoff is given by the finite
thickness $d$ of the disks, which put a natural cutoff of order $\ell\sim d$ to the growth of the logarithm. Despite this cutoff it is reasonable to expect that the behavior described above can be detected experimentally as a small $d$ amounts roughly speaking to increase $\ell$ by $d$ and as there is a quite large range of $\kappa=\ell/a$ where the repulsion is effective.

A second correction come from the non identity of the disks. This will be discussed in the next section, where it will shown that the difference in radii preserve the repulsive character of the force but for sufficiently small $\kappa$ transforms the logarithmic divergent force in a constant.

\section{Generalization\label{sezgen}}
It is natural to ask how the previous results generalize for an arbitrary system of two conductors.
The main lesson learned from the case of two equal disks is that it is better to exploit the finite limit
of the sum of capacity coefficients in writing the energy. In the general case we can take as basic quantities
\be \begin{split}
&C_{g1} = C_{11}+ C_{12}\,;\quad C_{g2} = C_{22}+ C_{12}\,;\\
&C = \dfrac{C_{11} C_{22} - C_{12}^2}{
C_{11}+ C_{22} + 2 C_{12}}
\end{split}
\label{varcap}\ee
In \cite{paf} we argued about the separate boundedness of $C_{g1}$ and $C_{g2}$. 
For $\kappa\to 0$ \eqref{intro1.6} imply 
\be C_T = C_{g1}(0) + C_{g2}(0)\label{ctcg1cg2}\ee

We note for the sequel that from the general relation
\eqref{intro1.2} if two conductors touch, acquiring the same potential, the total charge is redistributed on the two bodies with 
\be Q_a^V = \frac{C_{g1}(0)}{C_T}(Q_a+Q_b)\,;\quad Q_b^V = \frac{C_{g2}(0)}{C_T}(Q_a+Q_b)\,.\label{qvgen}
\ee
In terms of the variables \eqref{varcap} the energy takes the interesting form
\begin{align}
W &= \frac12 \dfrac{(Q_a+ Q_b)^2}{C_{g1}+ C_{g2}} + \frac12 \dfrac{ (C_{g2} Q_a - C_{g1} Q_b)^2}{(C_{g1}+ C_{g2})^2}\frac1{C}
\equiv \nonumber\\
&  \frac12 \dfrac{(Q_a+ Q_b)^2}{C_{g1}+ C_{g2}} + \frac12 
\left[\dfrac{C_{g1}(0)-C_{g2}(0)}{C_{g1}(0)+C_{g2}(0)}\frac{Q_a+Q_b}2 + \frac{Q_a-Q_b}2\right]^2\frac1{C}\label{energia1}
\end{align}
The second formulation being particularly useful for $Q_b=-Q_a$.
For $\kappa\to 0$ $C$ diverges then the energy acquire the expected value $W(0) = (Q_a+Q_b)^2/(2 C_T)$.

The form \eqref{energia1} shows clearly the crucial role for the forces of the dimension of the contact zone as $\kappa\to 0$.
The quantities $C_g$ 
behave like $A + B \kappa\log \kappa$ for $\kappa\to 0$, then, in the worst case, the first term in \eqref{energia1} can produce by derivation a
logarithmic singularities in the force. The second term depends strongly on the dimensionality. For point-like  contacts we expect the same behavior as sphere, i.e. $C\propto \log \kappa$. This imply that at short distances 
the leading term in the force is
\be 
F = -\dfrac{\partial}{\partial\ell}W \to \frac{1}2 \dfrac{ (C_{g2}(0)Q_a - C_{g1}(0) Q_b)^2}{C_T^2} \frac{1}{C^2}\dfrac{\partial}{\partial\ell} C 
 \propto \frac{1}{\log(\ell)^2}\,\frac1\ell
\label{forza1d}
\ee
As $C$ is a decreasing quantity with $\ell$, the force is attractive and divergent, the only exception being the vanishing of the coefficient of $1/C$ in \eqref{energia1} for $\ell\to 0$. This happens if the ratio of charges is the same as 
\eqref{qvgen}. One recognize in this analysis the direct generalization of Lekner work\cite{lek1} for spheres.
Spheres have a bonus: from the explicit computation of capacities\cite{maxw,lek1} it is easy to see that for $\kappa\to 0$
\be 
\begin{split}
&C_{g1} = - \frac{ab}{a+b} \left(\gamma_E + \psi(\frac b{a+b})\right) + {\cal O}(\ell) \\
&C_{g2} = - \frac{ab}{a+b} \left(\gamma_E + \psi(\frac a{a+b})\right) + {\cal O}(\ell) 
\end{split}
\label{cgspheres}
\ee
without logarithmic terms,
this means that these functions cannot produce forces for $\ell\to 0$, i.e. \eqref{forza1d} is the whole result for the second term in \eqref{energia1}. When the ratio of charges is the one in \eqref{qvgen} the first term in \eqref{energia1} produces, for spheres, the constant repulsive force
found by Lekner\cite{lek1} for this case.

The situation change completely if the contact zone is bidimensional, in this case $C\sim A/\ell$ for $\ell\to 0$ and the second term
in \eqref{energia1} produce a constant attractive force
\be F_2 = - \frac{1}2 \dfrac{ (C_{g2}(0)Q_a - C_{g1}(0) Q_b)^2}{C_T^2} \frac1A \label{forza2.1}\ee
again vanishing if the ratio of charges is the same as in \eqref{qvgen}. For $\ell\to 0$ then the form of the force is
\be F = F_2 + \frac{1}{2}\dfrac{(Q_a+Q_b)^2}{C_T^2} \frac{\partial}{\partial\ell}(C_{g1}+ C_{g2})
\label{forzagen}\ee

In this case the behaviour of the force for $\kappa\to 0$ rests on the subleading terms in $C_{g1}+C_{g2}$ for $\kappa\to 0$.
Very little is known on these terms in the general case. We have shown, cfr. eq\eqref{valorecgasint}, that in the case of two equal disks:
\[ C_{g1} = C_{g2} \to C_{g1}(0) + a \frac{\kappa}{2\pi^2}(\beta - \log(\kappa)) \]
giving rise to a logarithmic divergent force at short distances.

The case of two equal disks is  somewhat peculiar as we have only {\em one} linear combination at our disposal, $C_{11} + C_{12}$. In the general case we have two combinations, $C_{g1}$ and $C_{g2}$ and only the {\em sum} of these enters in the force for $\ell\to0$. The leading divergent terms cancel out in each combination, then one may wonder if additional cancellations arise when the two are combined. 

To study this problem we have considered a two disks capacitor with disks of {\em different} radii, $a, c$. We will use the variable $b=c/a$ and assume $c>a$. With this convention $b>1$. The rationale behind our hope for further cancellations is the following. In the general case we have two scales in the problem: $\kappa$ and $b-1$. The mathematical definition of the limit $\kappa\to 0$ is for $b-1$ fixed.
The logarithmic divergences for $\kappa\to 0$ can be smoothed by the fact that edge effects are distributed on a finite range, between $1$ and $b$ in the radial variable $r/a$. We study this problem partly analytically, using the results of ref.\cite{paf} and partly numerically. The results are given in the next section: we find only a {\em constant force} at short distances.

The consideration of $F_V$ and $F_E$ do not add anything new. In the case of constant potential from the general formula
\eqref{intro1.1} we have
\be F_V = \frac12 V^2 \frac{\partial}{\partial\ell} (C_{g1} + C_{g2}) \label{forzaVnew}\ee
For one earthed conductor the boundedness  of $C_{11}+ C_{12}$ as $\ell\to 0$ imply that the charge on the earthed conductor tends to 
$C_{12} Q_a/C_{11} \to - Q_a$ for $\ell\to 0$. This introduce a further depressive term in the second term of \eqref{forzagen}
leaving, for $\ell\to 0$
\be F_E = -\frac12 \dfrac{Q_a^2}{A}\,. \ee

\subsection{Numerical results for different disks}
The mathematical machinery for different disks is similar to the one already presented. The problem can be reduced\cite{paf} to the solution of a system of integral equations
\be
\begin{split}
&V_1 = F_1(t) +\int_0^b K(t,z;\kappa) F_2(z)\,dz\,;\\ 
&V_2 = F_2(t) +  \int_0^1 K(t,z;\kappa) F_1(z)\,dz\,.
\end{split}
\label{eqaccoppiatediv1}
\ee
with charges given by
\be Q_1 = \frac{2a}{\pi} \int_0^1 F_1(t)\,dt\,;\quad Q_2 = \frac{2a}{\pi} \int_0^b F_2(t)\,dt\,.\label{caichedef}\ee
The resolution of these equations allow the determination of capacitance coefficients.
Integrating equations \eqref{eqaccoppiatediv1} in the intervals $(0,1)$ and $(0,b)$ respectively and repeating the steps described for
\eqref{potcof2} we have the equivalent system
\begin{equation}
\begin{split}
\frac{\pi Q_1}{2a}  &= F_1(t) + \int_0^b K(t,s)F_2(s)ds- \int_0^b G(s;\kappa,1)) F_2(s)ds\\
\frac{\pi Q_2}{2a b}  &= F_2(t) + \int_0^1 K(t,s)F_1(s)ds -\frac1b\int_0^1 G(s;\kappa,b) F_1(s) ds\end{split}\label{potcof2div}
\end{equation}
which can be used for  a systematic study of densities. The function $G(t;\kappa,b)$ is a generalization of 
\eqref{defG1a}:
\be G(t;\kappa,b) = \frac1\pi\left[\arctan\frac{b-t}{\kappa} + \arctan\frac{b+t}{\kappa}\right]\label{defG1b}\ee
In ref\cite{paf} we have computed the values of all constants appearing in the force except the subleading term in $C_{g1}+ C_{g2}$.
\be
\begin{split}
C_{g1}(0) &= \frac{a}{\pi} \left( b - \sqrt{b^2-1}\right)\,;\qquad
C_{g2}(0) = \frac{a}{\pi} \left( b + \sqrt{b^2-1}\right)\\
C_T &= C_{g1}(0)+C_{g2}(0)  = \frac{2a}{\pi}b\,;\qquad
C = a\,\frac{1}{4\kappa} + \text{log terms}\,; 
\end{split}
\label{valoridischidiv}
\ee
The numerical computation of $C_{g1}$ and $C_{g2}$ gives a result parallel to the result for $C_g$ for single disks, i.e.
\be C_{g1} = C_{g1}(0) + \kappa ( b_1 + b_2 \log\kappa) \label{ris1}\ee
The results have the same quality of figure~\ref{figura1}, {\em but} the slopes $b_2$ have a different sign for $C_{g1}$ and $C_{g2}$, they in effect are {\em opposite} and cancel in the sum $C_{g1}+ C_{g2}$. This feature is most vividly shown if we plot
$\Delta C = (C_{g1}+C_{g2} - C_T)/\kappa$. As in figure~\ref{figura1} a possible logarithmic term must be represented by a straight line
in a linear-log plot.

\begin{figure}[!ht]
\begin{center}
\includegraphics[width=0.75\textwidth]{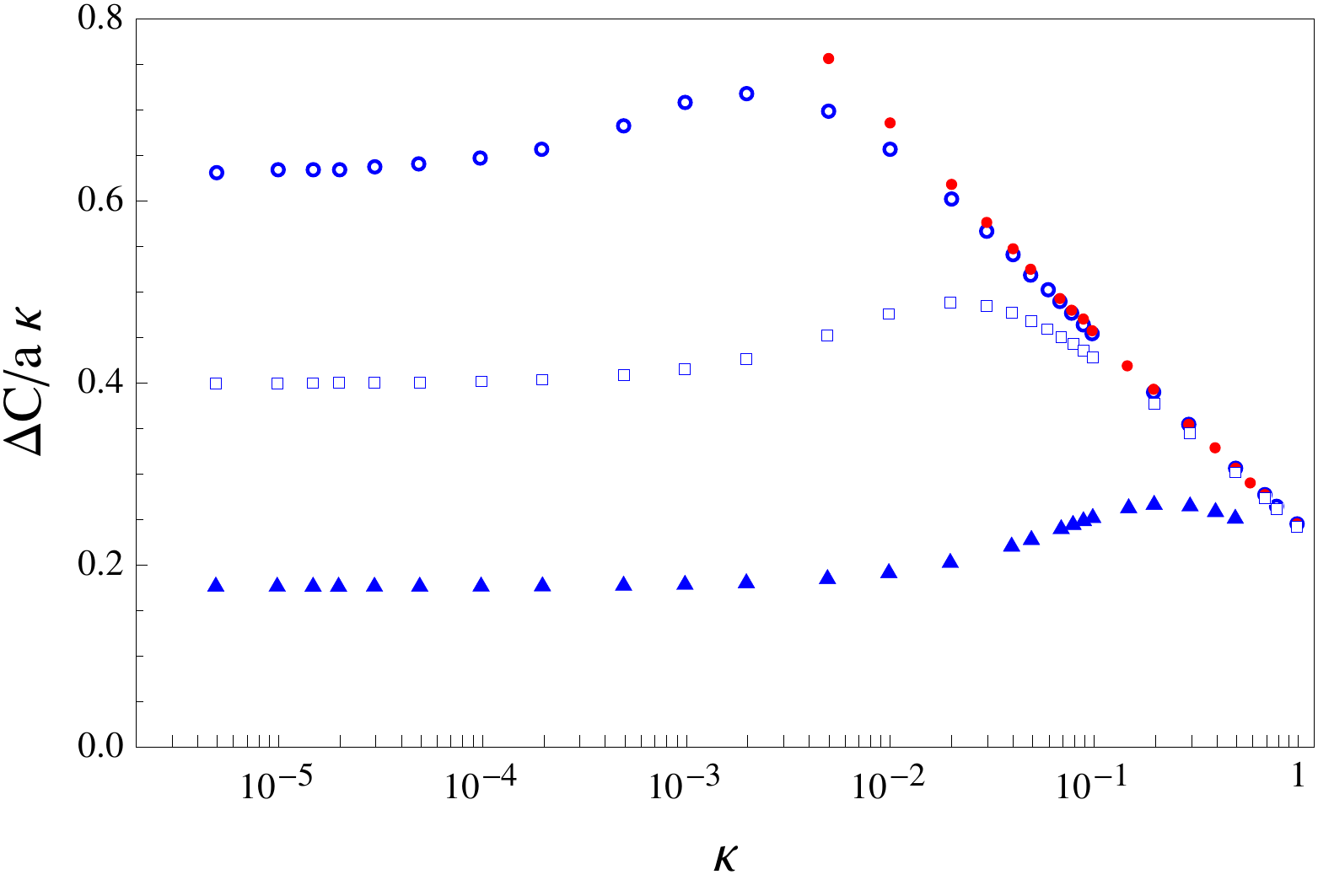}
\caption{The quantity $(C_{g1}+C_{g2} - C_T)/\kappa$ for different values of $b$, $b=1.1$, triangles,
$b=1.01$, open squares, $b=1.001$, open circles. The filled circles lying on a straight line are the results for $b=1$.\label{figura2LogV3}}
\end{center}
\end{figure}

Our results are summarized in figure~\ref{figura2LogV3} were we plot $\Delta C/\kappa$ for three different values of $b$: 1.1, 1.01, 1.001,
the lower points are for bigger $b$. A plateau is apparent for several order of magnitudes in $\kappa$, but the height of this plateau grows as $b\to 1$. In the figure it is plotted in the same units the result for $b=1$ (the straight line) which appear as the envelop of the curves with $b<1$.

In formulas the numerical computation indicates that for $\kappa\to 0$
\be C_{g1}+ C_{g2} = C_T + a B \kappa \label{valctot}\ee
The positive constant $B$ depends on $b$ and diverges for $b\to 1$. 
The constant $B$ is directly related to the force at constant potential, as from \eqref{forzaVnew} we have:
\[ F_V = \frac12 V^2 B\,. \]
Substituting  \eqref{valctot} and \eqref{valoridischidiv} in \eqref{energia1} and taking the derivative we obtain for the force
\be
 F = \frac1a\Bigl\{\frac{\pi^2}{8\pi^2 b^2} (Q_a+Q_b)^2 B - \frac1{2b^2} \left( b(Q_a-Q_b) + \sqrt{b^2-1}(Q_a+Q_b)\right)^2
 \Bigr\}
\ee
The force is constant, we cannot give a general rule for the sign as we do not explicitly know the constant $B$, but some general comments can be drawn. For $Q_b=-Q_a$ the force is attractive, a general result as we have discussed.

For large $b$ the first term vanishes and the second one gives an attractive limit force
\be F = -  2 \frac{Q_a^2}{a^2} \ee
The peculiarities of this case are evident: the force {\em does not depend} on the charge of larger disk. The constant in front of $Q_a^2/a^2$ is fixed. 

For $b\to 1$ instead the force become more and more repulsive. The enveloping features shown in figure \ref{figura2LogV3} 
can be translated in terms of potential energy:
the slope of the curve $U(\ell)$ for $\ell=0$  becomes bigger and bigger as $b\to 1$, in the limit case degenerate in a vertical tangent giving rise to the singularity for $b=1$.

\section{Conclusion}
In this work we study in a semi-analytical form the short distance limit of the force between two equal circular electrodes 
finding a repulsive logarithmic force in all cases except for two oppositely charged disks. We find that for disks with charges of different sign a point of null force. This point is present also for like charges when their ratio is sufficiently small.
The lesson learned by the solution of this problem suggests that the separation of the capacitance coefficients into terms of decreasing singularity can help in the general case. We apply this attitude to the general case of two conductors emphasizing the role of the dimension of the ``contact zone'' for the determination of the force. The surface contacts and the point-like contacts having  a completely different behavior. We checked semi-analytically our analysis by studying the force between two disks of different radii founding a constant force at short distances.


\end{document}